\documentstyle[12pt]{article}
\catcode`\@=11
\@addtoreset{equation}{section}

\newcommand{\numero}{SHEP 93/94-26}
\newcommand{\numerodeux}{McGill/94-24}
\newcommand{\Z}{{\cal Z}}
\newcommand{\F}{{\cal F}}

\newcommand{\beq}{\begin{equation}}
\newcommand{\beqn} {\begin{eqnarray}}

\newcommand{\eeq}{\end{equation}}
\newcommand{\eeqn} {\end{eqnarray}}

\begin{document}

\rightline{\numerodeux}
\rightline{\numero}
\vskip 2cm

\begin{center}
{\large THE QUARK PROPAGATOR\\ FROM THE DYSON-SCHWINGER
EQUATIONS:
\\ \  \\I. THE CHIRAL SOLUTION}
\vskip2cm
\renewcommand{\thefootnote}{*}
J.R.~Cudell\footnote{cudell@hep.physics.mcgill.ca},\\
{\small Department of Physics, McGill University}\\
{\small Montr\'eal, Qu\'ebec, H3A 2T8, Canada}\\
\ \\ \  \\
\renewcommand{\thefootnote}{\dag}
A.J. Gentles\footnote{ajg@hep.phys.soton.ac.uk}
\renewcommand{\thefootnote}{\ddag}
\ and D.A.
Ross\footnote{dar@phys.soton.ac.uk}\\
{\small Physics Department, The University}\\
{\small Southampton SO9 5NH, U.K.} \\
\ \\
June 1994
\renewcommand{\thefootnote}{*}
\vskip 2cm
{\bf Abstract}
\end{center}
Within the framework of the Dyson-Schwinger equations in the
axial gauge, we
study the effect that non-perturbative glue has on the quark
propagator.
We show that Ward-Takahashi identities, combined with the
requirement of
matching perturbative QCD at high momentum transfer, guarantee
the
multiplicative renormalisability of the answer. Technically, the
matching with
perturbation theory is accomplished by the introduction of a
transverse part
to the quark-gluon vertex. We show that this transverse vertex
is crucial for chiral symmetry breaking, and that massless
solutions exist
below a critical value of $\alpha_S$. Using the gluon  propagator
that
we previously obtained, we obtain small corrections to the  quark
propagator, which keeps a pole at the origin in the chiral phase.
\vfill\break
\section{Introduction}
The Dyson-Schwinger (DS) equations of motion are
one of the main tools for the  investigation of
non-perturba\-ti\-ve
effects. These equations are particularly suited to
the study of the interface between perturbative and
nonperturbative
regimes, as they allow us to compute the evolution of Green
functions when one enters the infrared region, and hence suggest
modifications to the usual Feynman rules.

In a previous paper, we have studied \cite{us} the DS equations
in the quenched approximation, {\it i.e.} for pure gauge QCD, and
found that, as was first suggested by Cornwall  \cite{cornwall},
some solutions for the gluon propagator are flatter than a  pole
in
the infrared region. In order to avoid the question of ghost
propagators, we worked in the axial gauge. Other groups
\cite{feynman,zw,review}  have also found that flatter solutions
exist, in covariant gauges,  although it is at present difficult
to
relate our results to theirs. The absence of a pole in the gluon
propagator at $k^2=0$ is natural if one assumes that gluons do
not
propagate to infinity, {\it i.e.} these solutions should describe
{\it confined} gluons. Furthermore, as was pointed out by
Landshoff and  Nachtmann \cite{LN}, the existence of such
solutions
is highly desirable in phenomenological
applications, and gives us a practical way to extend usual
perturbative
estimates to the  strongly-interacting sector of the theory. The
use
of these solutions has already met with some success in
diffractive
calculations \cite{pheno}.

The DS equations constitute an infinite tower of integral
relations
between n-point functions.  Hence, by themselves, they cannot be
solved. For instance, if we consider the DS equation for the
quark
two-point function (the propagator), it involves a three-point
function
(the quark-gluon vertex) as well as the gluon two-point function
(the
propagator). One can imagine solving the pure-gauge DS
equations for the gluon propagator as a first approximation,
hence we shall assume here that the gluon propagator is
known. However, resorting to higher-order DS equations to
calculate
the vertex will only bring in higher-order n-point functions,
which are
also a priori unknown. Hence one needs to use another input.

The simplest assumption is to take the perturbative $\gamma_\mu$
vertex, which is the so-called ``rainbow'' or ``ladder''
approximation.
One can be however more sophisticated, and follow the observation
of
Baker, Ball and Zachariasen (BBZ) \cite{bbz}, namely that one
needs
to choose a vertex that will obey the
Ward-Takahashi-Slavnov-Taylor
identities. These constitute another nonperturbative statement of
field
theory, and hence must be valid in general.

In our study of the gluon propagator \cite{us}, we chose for the
three-gluon vertex the simplest function that would obey the
Ward-Slavnov-Taylor identities. We showed that besides the
original
$1/k^4$  solution, there exists another solution, behaving like
$1/k^{0.2}$ near  $k^2=0$. To obtain a full picture of
nonperturbative
effects in QCD, one then  needs to calculate the quark propagator
corresponding to that gluon  propagator. As we shall see, the
problem
of multiple solutions arises again,  as for sufficiently small
$\alpha_S$, a chiral solution exists together with a  massive
one. We
shall show that one can constrain the spectral density of the
gluon
propagator and the value of the coupling so that only the massive
solution survives.

As the equation for the quarks is much simpler than that for the
gluons, we
investigate in more detail the effect of the ansatz for the
quark-gluon vertex on the quark propagator. Namely,  the minimal
vertex, used by BBZ for gluons and by Ball and Zachariasen (BZ)
for
quarks \cite{BZ}, leads to the breakdown of
multiplicative renormalisability.  We demonstrate that, following
the
method of Curtis and Pennington (CP) \cite{cp},
 it is possible to recover it by choosing a specific form for the
quark-gluon vertex, which agrees with perturbation theory in the
ultraviolet region, and which generalizes the B(B)Z ansatz. We
also
show that massive solutions exist only for the CP vertex. Simpler
ans\"atze lead to an inconsistent ultraviolet behaviour, unless
the
propagator is massless.  We also explain under which conditions
massless solutions will exist, even for the CP vertex.
We proceed to solve the DS equation in the massless phase, and
show
that nonperturbative  effects do not remove the pole of the quark
propagator in the chiral phase of  the QCD vacuum, suggesting
that
confined quarks can exist only after chiral  symmetry breaking.

This paper is set out as follows. In Section 2, we review our
results for the
gluon propagator, and outline the formalism that lead to these.
We point out
the problems linked with the BBZ choice of a three-gluon vertex.
In Section
3, we discuss the DS equation for the quark propagator in the
axial
gauge, and the approximations made when imposing the
Ward-Takahashi identity.
In Section 4, we consider the part of the propagator that
preserves
chiral symmetry. We show that the problem can be reduced to a
one-dimensional
integral equation, which we then renormalise. We show explicitly
that we
recover multiplicative renormalisability. In Section 5, we give a
criterion
for the existence of massless solutions, and then proceed to the
numerical solution of the equation in the massless case, for the
gluon propagator that we previously derived.
\section{The Gluon Propagator}
The fundamental observation \cite{bbz} is that one can use the
Ward-Slavnov-Taylor (WST) identities
to obtain a closed equation from the DS equation, if one neglects
quark loops and works  with gluons only. In the axial gauge ($n.A=0$),
where we do not have to  worry about ghost degrees of freedom,
the DS
equation relates the  propagator to the three- and four-point
vertices.

The three-point vertex, $\Gamma_\mu^{(3)}$ can be split into a
part
${\Gamma_T^{(3)}}_\mu$ transverse to the external gluon momentum
$k_\mu$,
${\Gamma_T^{(3)}}\cdot k=0$, and a longitudinal part,
${\Gamma_L^{(3)}}^\mu$, with ${\Gamma_L^{(3)}}\cdot k\neq 0$. As
the vertex is
a third rank tensor, depending on the three vectors entering the
vertex,  this split is not uniquely defined.
The WST identities relate ${\Gamma_L^{(3)}}$ to the propagator,
and these
identities can be ``solved'', hence producing an ansatz for the
longitudinal
part \cite{bbz,BZ}. It is this ansatz that we {\it define} as
being the
longitudinal part of the vertex in the following.

One can then obtain a closed equation for the propagator if one
makes two
assumptions. First, the full propagator is supposed to have the
same spin and
gauge structure as the free propagator:

\beq D_{\mu\nu}(k^2) \  = \ -  D_{\mu\nu}^0(k,n)
     \frac{\Z(k^2)}{k^2}  \label{gluprop} \eeq
where
\beq D_{\mu\nu}^0(k,n) \ = \  \delta_{\mu\nu}-\frac{k_\mu n_\nu +
n_\mu
k_\mu}{n.k}
                       + \frac{n.n \ k_\mu k_\nu}{(n.k)^2}
    \label{delta} \eeq
This spin structure projects out the four-point vertex
contribution to the
equation. If one further assumes that the three-point vertex is
dominated
by its longitudinal part ${\Gamma_L^{(3)}}$, which is itself
known from the
propagator via the WST identities, one obtains an equation which
involves
only the gluon propagator.

This equation still needs to be renormalised. One can show
without
approximation \cite{bbz} that a propagator with the spin
structure (\ref{gluprop}) has to be singular as $k^2\rightarrow
0$.
Thus the  inverse propagator should vanish at the origin. This
allows
one to subtract  the quadratic divergences, thus renormalising
the
gluon mass to zero and  cancelling the tadpole graphs. One
is then left with logarithmic divergences, which can be dealt
with through
wave function renormalisation at a renormalisation point $\mu_g$,
by defining
\begin{equation}
\Z(k^2)=\Z(\mu_g^2)\Z_R(k^2)\label{zrenorm}
\end{equation}
However, the resulting definition of the renormalised coupling constant
$\alpha_g$ is slightly different from the usual one:
\begin{equation}
\alpha_g(\mu_g)={\alpha_b \Z(\mu_g^2)\over 1+\alpha_b \Z(\mu_g^2)
\left[\int_b^\infty dk^2 \Psi(k^2,\Z)\right]_{\Z=\Z_R}}
\label{gcoupling}
\end{equation}
where $\alpha_b$ is the bare coupling and
$\Psi$ a function linear in ${\cal Z}$,
the explicit form of which can
be found in ref.~\cite{sch}. It has to be noted at this point
that
Eq.~(\ref{gcoupling}) leads to two related problems. First of
all,
the propagator asymptotically behaves like
$1/\log(k^2)^{11/16}$  as $k^2\rightarrow\infty$ \cite{bbz},
which is not what is obtained in perturbation theory.
Furthermore, one
obviously loses multiplicative renormalisability. This has the
nasty
consequence that, if the same truncation is
applied to the quark propagator, the
renormalised couplings are not equal anymore, because the
functions $\Psi$ are
different for quarks and gluons.

The solution to the resulting equation, found in
ref.~\cite{bbz}, has a very singular behaviour proportional to
$1/k^4$ as the
 momentum squared goes to zero. This was advertised  as a signal
for a
confining potential.
It has to be pointed out that in order to renormalise the
equation in the
presence of a $1/k^4$ singularity, one needs to
perform additional subtractions, and hence one obtains such
solutions only if
one assumes they exist.
In ref.~\cite{us}, we exploited the fact that the
 integral equation obtained in  ref.~\cite{bbz} is non-linear and
can
in  general admit several solutions. We solved the
original  equation, without additional subtractions, and obtained
a
gluon propagator which, although singular as $k^2 \rightarrow 0$
 as required in the axial gauge,
has a cut singularity,
 flatter than a pole. This describes a soft gluon which is
confined
rather than
 confining and can conveniently be used in models of the pomeron
such
as
 that suggested by Landshoff and Nachtmann \cite{LN}. This
propagator may
be written in the axial gauge ($n.A=0$) as
\beq {Z_R(k^2)\over k^2} \ ={\mu_g^{-2}\over \ a_1 \left(
\frac{k^2}{\mu_g^2} \right) ^ {b_1}
                  + a_2 \left( \frac{k^2}{\mu_g^2} \right) ^
{b_2}
                  +c\ \ln \left[ d \left(\frac{k^2}{\mu_g^2}
\right)+ e \right]}
    \label{gluz} \eeq

\noindent The dimensionless constants
$a_1,b_1,a_2,b_2,c,d$ and $e$, are given in the
Table. The mass scale $\mu_g$ is the value of momentum at which
$\Z_R(k^2)$ takes the value 1, so that the propagator matches the
value
 of the free gluon propagator. It is not obtained from the DS
equation,
which is scale invariant, but rather from a fit to the total
cross-section for proton-proton scattering calculated using this
propagator. An optimum value of 0.8~GeV was found for $\mu_g$. It
is
worth pointing out that most of the ideas discussed here can be
applied to other theories, living at other mass scales, such as
technicolour, and that one would only need to adjust $\mu_g$ to
do so.

\begin{center}
{\bf Table:} The constants used in Eq.~(\ref{gluz})  \end{center}

\begin{center}
\begin{tabular}{|c|c|c|c|} \hline
 $a_1$ & 0.88 &$c$   & 0.59\\ \hline
 $a_2$ & -0.95& $d$   &  2.1 \\ \hline
 $b_1$ & 0.22& $e$   &  4.1 \\ \hline
 $b_2$ & 0.86& $\alpha_g(\mu_g)$& 1.4\\  \hline
\end{tabular}
\end{center}

Note that in the case of a multiplicatively renormalisable
theory, the actual
value of $\alpha_S$ should not matter.
In principle, one could always rescale $\alpha_S$ by varying
$\mu$ according to
\begin{eqnarray}
\alpha_S(\mu'){\cal Z}_R(\mu'^2,q^2)&=&\alpha_S(\mu){\cal
Z}_R(\mu^2,q^2)\nonumber\\ {\cal Z}_R(\mu^2,q^2)&=&{\cal
Z}_R(\mu^2,\mu'^2) {\cal Z}_R(\mu'^2,q^2) \label{RG}
\end{eqnarray}
where the first argument of ${\cal Z}$ refers to the
renormalisation scale:
${\cal Z}(\mu^2,\mu^2)=1$. However, our gluonic solution does not
obey
these  equations because of (\ref{gcoupling}), and one obtains a
\renewcommand{\thefootnote}{1}
solution only for a  given value\footnote{Note that
there was a mistake of a  factor 2 in our definition of
$\alpha_g$ in
Refs.~\cite{us}}. This value of $\alpha_g(\mu_g)$
enters phenomenological calculations, which then determine
$\mu_g$.

Hence the assumption of a given spin structure (\ref{gluprop})
and the neglect
of the transverse part of the three gluon vertex leads to a DS
equation which admits two solutions, and which breaks
multiplicative
renormalisability. In order
to explain the discrepancy of the BBZ result with more recent
ones, the
assumed spin structure of the gluon propagator has been put into
question \cite{review}. However,  there is no doubt that
such solutions exist. Further solutions can of course
exist, given that the equation is highly nonlinear. The neglect
of the
transverse part of the vertex, on the other hand, may have
important
consequences, which we shall explore in this paper for the quark
propagator.
\section{The Dyson-Schwinger equation for quarks}
Following their work with Baker on the gluon propagator
\cite{bbz},
Ball and  Zachariasen have considered the DS equation for the
quark
propagator \cite{BZ}. Choosing again a purely longitudinal
quark-gluon
vertex,  they found that the quark propagator
corresponding to a $1/k^4$ gluon  was suppressed, and became
constant
near the origin. Furthermore, they  showed that a chiral solution
was
always possible, and argued that a solution
could be found which would break chiral symmetry.
We shall see that, in the case of a less singular
gluon propagator, only the chiral solution is possible if one
neglects
the transverse part of  the vertex. However, implementing the
improvements proposed by Curtis and Pennington  \cite{cp}, we
shall
show that multiplicative renormalisability (and hence the
transverse
part of the vertex) leads to an equation that allows chiral
symmetry
breaking, and even suggests a range of parameters (describing the
coupling and the gluon propagator) which would lead to a unique,
massive, solution.
\subsection{The Ball-Zachariasen equation}
We define the propagator for a quark with momentum $q$ as:
\beq S(q) \ = \ F(q^2) \gamma \cdot q + G(q^2)
       \label{quarkprop} \eeq
so that $F(q^2)$ represents the chiral-symmetry conserving part
and $G(q^2)$
represents the chiral-symmetry breaking part. We shall also use
the equivalent
notation:
\beq S(q) \ = {{\cal F}(q^2) \left[\gamma \cdot q +
\Sigma(q^2)\right]\over
q^2-\Sigma^2(q^2)}
       \label{quarkpropb} \eeq

The free propagator is therefore
obtained by setting $F(q^2)$ to $1/q^2$ and  $G(q^2)$ to zero
(neglecting  all current masses). In the axial gauge
Eq.~(\ref{quarkprop}) is not
the most general form for the quark propagator. First of all, the
functions $F$ and $G$ can depend on $n.q$ as well as $q^2$. As in
the case of the gluon propagator \cite{bbz,us}, we seek solutions
for which these functions are independent of $n.q$. Furthermore,
the gauge dependence can also arise through extra spin
structures, proportional to $\gamma \cdot n$ and $n.q$, which
although absent for the free propagator can
in general occur for the dressed propagator. As pointed out in
ref.~\cite{BZ}), these extra terms drop out of the equations for
$F$ and $G$ if one specializes to a gauge vector orthogonal to
$q$. This gauge choice $n.q=0$ is the
only one that makes the algebra sufficiently tractable that
reliable solutions to the DS equations can be obtained.
The question of the dependence of the quark propagator on $n$,
although important, is difficult to address in the axial gauge, and beyond the
scope of this paper.

  The DS equation for $S(q)$,
sketched in Fig.~1, in Euclidean space is given by
\beq 1 \ = \gamma \cdot q\  S_b(q) \  - \ C_F\ {\alpha_b}
    \int \frac{d^4k}{4\pi^3\ k^2} \Z_b(k^2) \gamma^\mu D_{\mu
\nu}(k,n)
                   {S}_{b}(q-k) \Gamma^\nu(q-k,q) {S}_{b}(q)
\label{Dyson}\eeq
where the subscript $b$ on $\alpha$, $\Z$ and $S$ indicate that
these are bare
quantities which will have to be renormalised. $C_F=4/3$ is the
quark
Casimir invariant.

\begin{figure}
{\small \noindent Figure 1: A pictorial representation of
Eq.~(\ref{Dyson}).
The hatched circles represent the exact two-point functions, and the
cross-hatched circle the exact
three-point function. }
\end{figure}

In general it is not possible to solve this equation because it
involves the
unknown quark-gluon vertex function $\Gamma^\nu(q-k,q)$. This
vertex function is related to a four-point Green function via a
DS
equation. Thus we get an infinite tower of coupled integral
equations. On
the
other hand, in the axial gauge, this quark-gluon vertex function
connecting a
quark of momentum $p\equiv q-k$ to one of momentum $q$ through
the absorption of a
gluon of momentum $k$, obeys the Ward-Takahashi (WT) identity
sketched in
Fig.~2:
\beq k^\nu \Gamma_\nu(p,q) \ = \ S(q) ^{-1}\ -\ S(p)
^{-1}\label{ward} \eeq

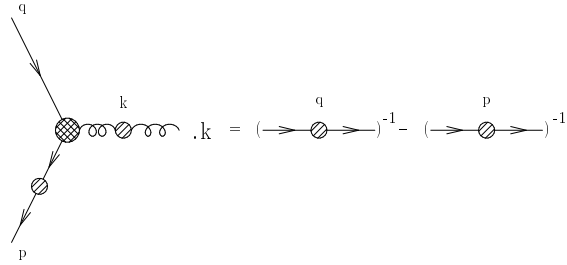
{\small \noindent Figure 2:  A representation of the Ward-Takahashi identity of
Eq.~(\ref{ward})
using the same convention as for Fig.~1.}

One can then solve this equation to determine  $\Gamma^\mu_L$,
the part of the vertex
longitudinal to $k$. Following ref.~\cite{BZ}, we obtain:
\begin{eqnarray} -S(q)\Gamma_\mu^L(p,q) S(p)&=& {1\over 2}
\left[F_b(p)+F_b(q)\right]\gamma^\mu +{1\over 2}
\left[F_b(p)-F_b(q)\right]{2\gamma\cdot q\ \gamma^\mu\
\gamma\cdot
p\over p^2-q^2}\nonumber\\
&+&{1\over 2}
\left[F_b(p)-F_b(q)\right]{q^2+p^2\over
p^2-q^2}\ \gamma^\mu\nonumber\\
&+&\left[G_b(p)-G_b(q)\right]{\gamma^\mu\ \gamma\cdot
p+\gamma\cdot q\ \gamma^\mu\over p^2-q^2}\label{BZvert}
\end{eqnarray}
The simplest approach consists, as suggested in ref.~\cite{DW},
in
neglecting the transverse part of the vertex: one then
assumes that the integral equation is dominated by the part
(\ref{BZvert}) which is
\renewcommand{\thefootnote}{2}
determined from the above WT identity.\footnote{In other gauges
the
Ward identity Eq.~(\ref{ward}) must be replaced by the full
Slavnov-Taylor identities \cite{ST} which involve the
interactions of
Faddeev-Popov ghosts. Nevertheless it was argued in
ref.~\cite{mike1}
that the contributions from Faddeev-Popov ghosts are small and so
the
Ward identity of Eq.~(\ref{ward}) was imposed on the integral
equation
formulated in a covariant gauge.} The resulting  integral
equations
for the two functions $F_b(q^2)$ and $G_b(q^2)$ separate and we
have
\begin{eqnarray}
 1 \ & = & q^2\ F_b(q^2) -C_F\ {\alpha_b} \int \frac{d^4k}{4\pi^3
\ k^2}
   \Z_b(k^2) D_{\mu \nu}(k,n)
     \left\{ \frac{F_b(q^2)+F_b(p^2)}{2}\delta_{\mu\nu} \right.
\nonumber \\
         &+& \left.   \frac{1}{(q^2-p^2)}
   \left[  F_b(q^2)-F_b(p^2) \right]
 \left( \frac{1}{2}k^2 \delta_{\mu\nu}+q_\mu p_\nu + p_\mu q_\nu
 \right) \right\}   \label{feqa} \end{eqnarray}
and
\begin{eqnarray}
 0 \ & = &\ G_b(q^2) -C_F\ {\alpha_b} \int \frac{d^4k}{4\pi^3 \
k^2}
   \Z_b(k^2) D_{\mu \nu}(k,n)
      \nonumber \\
     & &  \hspace{-.5in} \left\{    \frac{1}{(q^2-p^2)}
    \left[ G_b(q^2)-G_b(p^2) \right]
 \left( q \cdot k \delta_{\mu\nu}+q_\mu p_\nu + p_\mu q_\nu
 \right) \right\}   \label{geq} \end{eqnarray}
 where again we have introduced the subscript $b$ on $F$ and $G$
to
indicate that these quantities are to be renormalised. (There is
no
renormalisation for $\Sigma$ since this function is zero for the
free
propagator, i.e. in the absence of an explicit current mass there
is
no parameter in the QCD Lagrangian to renormalise, hence the
ratio
$F/G$ does not get renormalised).
\subsection{Multiplicative renormalisability and the transverse
part of the
vertex}
As was observed in \cite{BZ}, Eq.~(\ref{feqa}, \ref{geq}) are not
multiplicatively renormalisable. In other words, if we
renormalise the
quark and gluon wavefunctions by imposing $F_b(q^2) =
F_b(\mu_f^2)$
$F_R(q^2)$, $G_b(q^2) = F_b(\mu_f^2)$
$G_R(q^2)$ and  $\Z_b(q^2)=\Z_b(\mu_g^2) \Z_R(q^2)$, additional
terms
need to be introduced in the  definition of the renormalised
coupling,
as in Eq.~(\ref{gcoupling}).  These terms are not the same in the
quark and in the gluon case, so one loses  not only
multiplicative
renormalisability, but also the universality of the  QCD coupling
constant!

These problems come from the ultraviolet region, and can be
traced back
to the neglect of the transverse part of the  vertex in the
solution of
the WT identities: the vertex (\ref{BZvert}) does not match the
perturbative one at high momentum transfer. Hence, it is
necessary to
postulate a transverse part that will restore multiplicative
renormalisability. Curtis  and Pennington have shown \cite{cp}
that
this goal can be achieved in QED by  considering the perturbative
limit
of the vertex. Their argument can be trivially extended to QCD in
the axial gauge: the one-loop corrections to the propagator are
identical to those of QED up to the quark Casimir invariant
$C_F$, and the
vertex corrections are also the same: because the vertex and the
wavefunction renormalisation constants $Z_1=Z_2$ are equal in
this
gauge, and because $Z_2$ is a function of $C_F$ only, the
diagrams
involving the three-gluon vertex, which depend on $C_F-C_A/2$,
with $C_A$ the
adjoint Casimir invariant, have to
cancel. Hence the argument is totally similar to that of
ref.~\cite{cp} and goes as follows.

In general, the quark propagator has the perturbative
limit \beq \lim_{q^2\rightarrow \infty}\F(q^2)=1+{\alpha_S\
\xi\over  4
\pi}\ln{q^2\over\Lambda^2}\label{pertf}\eeq with $\xi$ the
anomalous
dimension. The one-loop vertex can be shown \cite{cp} to tend to
\begin{eqnarray} \lim_{q^2/p^2\rightarrow \infty}
\Gamma_{\mu}^{pert}(p,q)&=&\gamma_\mu\left[1- {\alpha_S\ \xi\over
4\pi} \ln\left({q^2\over\Lambda^2}\right)\right]\nonumber\\
&-&{\alpha_S\over 4\pi} {\ln\left({q^2\over\Lambda^2}\right)\over
q^2}
(\gamma\cdot p\ \gamma_\mu\gamma\cdot p-q_\mu\gamma\cdot p+\xi
q_\mu\gamma\cdot p) \label{gammapert}\end{eqnarray}

One can compare this answer with the vertex
(\ref{BZvert}) in the same limit, to conclude
that the transverse part must give:
\beq \lim_{q^2/p^2\rightarrow \infty}\Gamma_\mu^T(p,q) =
{\alpha_S\ \xi
\ln\left({q^2\over\Lambda^2}\right) \over 4\pi q^2}
(-q_\mu\gamma\cdot p
+\gamma\cdot q\gamma_\mu\gamma\cdot p)\label{transversepert}\eeq
This tensor is indeed transverse to $q\approx k$.
One can then extend the tensor structure of this vertex so that
it becomes transverse to $k$ for any value of $p$ and $q$.
We find that the simplest extension is:
\beq S(p) \Gamma_\mu^T(p,q) S(q)=
\left(\F(q)-\F(p)\right) {\gamma_\mu
(q^2-p^2)-(q_\mu+p_\mu)(\gamma\cdot q -\gamma\cdot p)\over
{\cal D}}\label{transverse}\eeq
where ${\cal D}$ is an expression symmetric in $p$ and $q$, and
behaving as $q^4$ in the large-$q^2$
limit. ${\cal D}$ must not introduce any singularity, must be
symmetric in  $q$ and $p$, an must satisfy $\Gamma_\mu^T(p,p)=0$.
In general,
\beq
{\cal D}= (q^2+p^2)^2 \ \eta\left({q.p\over (q^2+p^2)},{q^2 p^2\over
(q^2+p^2)^2}\right)
\label{dval}
\eeq
with $\eta$ a regular function such that $\eta(0,0)=1$.

It is worth noting that the tensor structure of the transverse
vertex (\ref{transverse}) can be generalised. Indeed, Ball and
Chiu \cite{BC} have given a set of 8 independent tensor
structures spanning the space of regular transverse vertices.
Only three of these have the correct helicity structure to
contribute to the DS equation in the chiral limit, and are (in
the notation of Ball and Chiu):
\begin{eqnarray}
T_2^\mu&=&[p^\mu (k.q)-q^\mu (p.k)] \gamma\cdot(p+q)\nonumber\\
T_3^\mu&=&k^2\gamma^\mu-k^\mu\gamma\cdot k\nonumber\\
T_6^\mu&=&\gamma^\mu (q^2-p^2)-(q+p)^\mu \gamma\cdot (q-p)
\end{eqnarray}

The transverse vertex given by Eq.~(\ref{transverse}) corresponds
to:
\beq
\Gamma^T(p,q)=
\frac{1}{\cal D} \left( \frac{1}{{\cal F}(q)} - \frac{1}{{{\cal F}(p)}}\right)
  \left(
{1\over 2}(p^2+q^2) T_6 + {1\over 2}(p^2-q^2) T_3
+...\right )
\eeq
where the ellipses refer to terms that vanish in the DS equation.
This differs from the structure of ref.~\cite{cp}, where the
transverse vertex was chosen to be proportional to $T_6$. This
can be understood from the fact that the expression of the
longitudinal part of the vertex used here differs from theirs, by
a transverse tensor proportional to $T_3$, in such a way that we
get the
same high-$q^2$ or $p^2$ leading term for the total vertex.  The
addition of such extra subleading terms is always possible, and
we shall investigate the effect of these through
the variation of function $\cal D$ in Eq.~(\ref{transverse}).

The transverse vertex (\ref{transverse}), together with the
longitudinal one (\ref{BZvert}), can then be included in
Eq.~(\ref{Dyson}) to obtain:
\begin{eqnarray}
 1 &=&  p^2 F_b(p)+ C_F\ {\alpha_b} \int
{d^4k\over 4\pi^3 k^2} D^b_{\mu\nu} \Z_b(k^2)\nonumber\\
&\biggl\{&-{1\over 2} \left[ F_b(p)+F_b(q)\right]\delta_{\mu\nu}
-{F_b(p)-F_b(q)\over p^2-q^2}
\left( p_\nu q_\mu+p_\mu q_\nu +{1\over 2} (p-q)
\delta_{\mu\nu}\right) \nonumber\\
&+&{\left(p^2 F_b(p)-q^2 F_b(q)\right)\left(\delta_{\mu\nu}
(q^2-p^2)+p_\mu p_\nu -q_\mu
q_\nu +p_\mu q_\nu  - p_\nu q_\mu\right)\over {\cal D}}\biggr\}\
\
\label{feq} \end{eqnarray}
\section{Solving the equation}
\subsection{The angular integral}
The angular integral in Eqs.~(\ref{feqa}, \ref{geq}, \ref{feq})
cannot
be performed analytically without prior knowledge of the function
$F$,
since this occurs with the argument $(q-k)^2$. The approximation,
first proposed by Schoenmaker \cite{sch}  in the gluon case,
consists
in replacing a function $f((q-k)^2)$ by  $f(q^2+k^2)$. This is
clearly
valid in the regions $k^2>>q^2$ and $k^2<<q^2$.  One then needs
to
find a function $f$ which is sufficiently slowly varying  that
the
error generated by the integration over the whole range of $k^2$
be
negligible.

The first choice would be $f=F$. In this case, the contribution
of the
longitudinal part of the quark-gluon vertex to the DS equation
vanishes. Hence the BZ equations (\ref{feqa}, \ref{geq}) admit
only the
trivial solution $F(q^2)=1/q^2$, $G=0$. If the transverse part of
the
vertex is added, the resulting equation (\ref{feq}) leads to a
propagator  close to the perturbative one, with a pole at
$k^2=0$.
Neither solution is a slowly  varying function, hence the
approximation is not justified in this case.

In the following, we shall use $f={\cal F}$.
Our solutions show that this assumption is reasonable
so that the above ansatz is justified {\it a posteriori}.
 The only region where this approximation may have led to
substantial
errors would be for small $q^2\approx k^2$, but then the
integrand is suppressed because (as can be seen from
Eq.~(\ref{gluz}))
the function $\Z_R(k^2)$ vanishes. This approximation keeps the
contributions both  from the longitudinal and from the transverse
part
of the vertex, hence we shall obtain nontrivial solutions both
for
Eqs.~(\ref{feqa}, \ref{geq}) and (\ref{feq}).

With this approximation the angular part of the integration over
momentum $k$ in Eqs.~(\ref{feqa}, \ref{geq}, \ref{feq}) may be
performed, and both equations can be recast into the following
form if
$G=0$: \begin{eqnarray} 1 \ &=&\ q^2 F_b(q^2) \nonumber\\ &-& \
{C_F\
{\alpha_b}\over 4\pi}
 \int dk^2 \left[ {\cal Z}(k^2) \Delta_1(k^2/q^2) \ F_b(q^2) + \
 {\cal Z}(k^2) \Delta_2(k^2/q^2) F_b(k^2+q^2) \right]
\label{equation}
    \end{eqnarray}
The kernels $\Delta_1$ and $\Delta_2$ are given, both in the case
of
Eqs.~(\ref{feqa}, \ref{geq}) and in the case of Eq.~(\ref{feq})
in the
Appendix. This is a Fredholm equation of the second
kind which must now be solved numerically.
Before we can do this, however, we must consider the question of
the
renormalisations required to absorb the ultraviolet divergences.
\subsection{Renormalisation}
The renormalisation procedure is similar whether one neglects the
transverse part of the vertex or not. The renormalisation
constant
${\cal Z}_b(\mu_g^2)$ has been introduced in
Eq.~(\ref{zrenorm}). As mentioned in Section 2 we choose this
constant such that
the renormalised function, $\Z_R(q^2)$ takes the value 1
at $q^2=\mu_g^2$, with $\mu_g$
taken to be 0.8 GeV.  We likewise introduce a renormalisation
constant
$\F_b(\mu_f)$ such that $$ F_R(q^2) \ =
{F_b(q^2)\over\F_b(\mu_f)}  $$ and
$$G_R(q^2) \ = {G_b (q^2)\over \F_b(\mu_f)} $$ are ultraviolet
finite
and we choose it
so that the quantity $\F_R(q^2)$ takes the value 1 at $q^2 =
\mu_f^2$.
Eq.~(\ref{equation}) then becomes finite if one rewrites it in
terms of the coupling
\beq \alpha_f \ =
\ \frac{\alpha_b \Z(\mu_f)}{1-\alpha_b {\cal Z}(\mu_f) C_F/4\pi
\int dk^2 {\cal
Z}(k^2) \Delta_1(k^2,\mu_f^2)}  \label{newcoupling}
\eeq
and the renormalised equation (\ref{equation}) becomes:
\begin{eqnarray}
1 &=& \left( 1-{C_F\ \alpha_f\over 4\pi} \int dk^2 {\cal
Z}_R(k^2)
 \left[ \Delta_1(k^2,q^2) - \Delta_1(k^2,\mu_f^2) \right] \right)
F_R(q^2)
 \nonumber \\
&-&{C_F\ \alpha_f\over 4\pi} \int dk^2 \Z_R(k^2)
\left(\Delta_2(k^2,q^2)F_R(k^2+q^2)
   - \Delta_2(k^2,\mu_f^2) F_R(k^2+\mu_f^2)\right) \label{fred0}
\end{eqnarray}
\subsubsection{Renormalised BZ equation}
In the case of the BZ kernels, the UV divergence of the equation
is
concentrated in the $\Delta_1$ term, which has a log singularity
as
$k^2\rightarrow\infty$. This means that the coupling
(\ref{newcoupling}) cannot be related to the usual renormalised
QCD
coupling \beq\alpha_S(\mu_f)=Z_b(\mu_f)\alpha_b\label{alphas}\eeq
Hence one loses the universality of the QCD coupling.

Furthermore, the equation for the chirality breaking term $G$ is
identical to Eq.~(\ref{fred0}) with the replacement $F\rightarrow
G$,
$1\rightarrow 0$. The leading behaviour at high $q^2$ comes from
the
remnant of the UV term. If we assume, in agreement with the
renormalisation group, that ${\cal Z}_R(q^2)\sim 1/[q^2\log(q^2)]$
as
$q^2\rightarrow\infty$, the leading terms of the equation for the
$G$
term are:
\beq 0   \approx G_R(q^2)-{C_F\ \alpha_f\over
4\pi}\log(\log(q^2)) G_R(q^2) \eeq
 which has only $G=0$ as a consistent
solution. Hence, unless the gluon propagator is highly singular,
the
BZ equation does not lead to chiral symmetry breaking.

If one overlooks these problems, and goes ahead to solve the
equation, one
obtains a function $F_R(q^2)$ which is very close to the
perturbative $1/q^2$.
The propagator in this case keeps a pole, the residue of which is
slightly
bigger than the perturbative one.

As we shall see, the chiral solution will continue to possess
these
properties, even after we get a consistent equation for the quark
propagator
by the introduction of the transverse part of the vertex.
\subsubsection{A multiplicatively renormalised equation}
The main effect of the inclusion of the transverse vertex is to
shift the divergence from the integral of $\Delta_1$  to
that of $\Delta_2 F$ in Eq.~(\ref{fred0}).
Indeed, the integral over $k^2$ of $\Delta_2$ in
Eq.~(\ref{equation}) now diverges logarithmically since
$\Delta_2(k^2,q^2)$ behaves like $1/k^2$ as $k^2 \rightarrow
\infty$, whereas
the integral of the $\Delta_1$ term is finite.

This fact means that the renormalised coupling
(\ref{newcoupling}) is related
to the usual one (\ref{alphas}) by a finite renormalisation:
\beq
\alpha_f={\alpha_S(\mu_f)\over
1-\alpha_S (\mu_f) C_F/4\pi \int dk^2 {\cal Z}(k^2)
\Delta_1(k^2,\mu_f^2)}  \label{couplingrelation}
\eeq

Hence we can now write our equation in terms of the true QCD
coupling:
\begin{eqnarray} {1 \over \F_b(\mu_f)} &=&\ q^2 F_R(q^2)
\nonumber\\
&-& \ {C_F\ \alpha_S(\mu_f)\over 4\pi}
 \int dk^2 \left[ {\cal Z}_R(k^2) \Delta_1(k^2,q^2) \ F_R(q^2)
\right.\nonumber\\
&+&\left. \
 {\cal Z}_R(k^2) \Delta_2(k^2,q^2) F_R(k^2+q^2) \right]
\end{eqnarray}

Setting $q^2=\mu_f^2$ in the above, we obtain:
\beq {1 \over \F_b(\mu_f)} = 1  - \ {C_F\ \alpha_S(\mu_f)\over 4\pi}
 \int dk^2 \left[ {\cal Z}_R(k^2) {\Delta_1(k^2,\mu_f^2)\over
\mu_f^2}  + \
 {\cal Z}_R(k^2) \Delta_2(k^2,\mu_f^2) F_R(k^2+\mu_f^2) \right]
\eeq

We can then equate both expressions for $1/\F_b(\mu_f)$ to
obtain:
\begin{eqnarray}
1 &=& \ q^2 F_R(q^2)\nonumber\\
 &-& \ {C_F\ \alpha_S(\mu_f)\over 4\pi}
 \int dk^2 \left[ {\cal Z}_R(k^2) \Delta_1(k^2,q^2) \ F_R(q^2)
- {\cal Z}_R(k^2) {\Delta_1(k^2,\mu_f^2)\over \mu_f^2}
\right]\nonumber\\
&-&{C_F\ \alpha_S(\mu_f)\over 4\pi} \int dk^2
 {\cal Z}_R(k^2)\left[\Delta_2(k^2,q^2) F_R(k^2+q^2) -
 \Delta_2(k^2,\mu_f^2) F_R(k^2+\mu_f^2) \right] \label{fred}
\end{eqnarray}
As can be seen the large $k^2$ behaviour of ${\Delta}_2$ cancels
out and
the integral is now ultraviolet convergent.
\section{The chiral solution}
\subsection{Asymptotic behaviour}
The first test of the consistency of our
results is a direct comparison of equation
(\ref{fred}) with perturbation theory, {\it i.e.} for
$q^2\rightarrow\infty$.  As we have explained, before
subtractions,
the equation is ultraviolet  divergent because of the terms
proportional to $\Delta_2$. After subtraction,  the leading
$\log{q^2}$ comes from those terms in $\Delta_2$ that behave like
$1/k^2$. The equation then becomes, for $q^2\rightarrow\infty$:
\beq
1 \approx \ q^2 F_R(q^2)
-{C_F\ \alpha_S\over 4\pi} \int_{\mu_f^2}^{q^2} {3 dk^2\over 2
k^2}
F_R(k^2)\left[
 {\cal Z}_R(k^2)
\right] \label{fredasym}
\eeq
Writing $q^2 F_R(q^2)\approx 1+\xi \log(q^2)$, $Z_R\approx 1$,
one then gets a consistent solution to order $\xi\sim\alpha_S$,
provided that
\beq
\xi={3 C_F\ \alpha_S\over 8\pi}\label{anomalous}
\eeq
in agreement with one-loop results (in axial gauge).

Furthermore, the equation also agrees with RG-improved
perturbation theory.
Assuming that for large $q^2$ one has $Z_R(q^2)\approx
\log(q^2)^{-1}$,
one then gets a consistent asymptotic behaviour $q^2 F_R(q^2)
\sim C+\log(q^2)^{-\xi}$,
with $C$ a constant, and $\xi$ still given by
Eq.~(\ref{anomalous}).
Hence we see that Eq.~(\ref{fred}) encompasses our perturbative
knowledge of
propagators.

Finally, as we shall now explain, it suggests that the $G=0$
solution
cannot be valid for arbitrary values of $\alpha_S(\mu_f)$.
\subsection{Critical value of $\alpha_S$ and chiral symmetry
breaking}
By shifting the $k^2$ integration in the terms proportional to
$\Delta_2$, one
can recast the equation in the following form:
\begin{equation}
\phi(q^2) \F_R(q^2)=\phi(\mu_f^2)+\int dk^2 {\cal K}(q^2,k^2)
\F_R(k^2)\label{fredholm}
\end{equation}
with:
\begin{eqnarray}
\phi(q^2)&=& 1 - \ {C_F\ \alpha_S\over 4\pi}
 \int dk^2 \left[ {\cal Z}_R(k^2) {\Delta_1(k^2,q^2)\over q^2}
\right]\nonumber\\
{\cal K}(q^2,k^2)&=&{C_F\ \alpha_S\over 4\pi}{1\over k^2}\left[
 {\cal Z}_R(k^2-q^2)\Delta_2(k^2,q^2) -
 {\cal Z}_R(k^2-\mu_f^2) \Delta_2(k^2,\mu_f^2)  \right]
\nonumber\\
\end{eqnarray}

In the case where $\phi(q^2)\neq 0$ for all $q^2$, the equation
can be reduced to
a Fredholm equation of the second kind. Its kernel ${\cal K}$
is integrable
and bounded, so that there is a unique solution. We obtain it by
discretising
Eq.~(\ref{fredholm}) and inverting the matrix equation thus
obtained, which is
equivalent to Fredholm's solution \cite{Fredholm}. We then
get a smoother solution by introducing the obtained points as the
input of an
iterative method, where we use the left-hand side of
Eq.~(\ref{fredholm}) as
the output, and the left-hand side as the input, which is the
Liouville-Neumann method \cite{Liouville}. This converges nicely
as
long  as $\phi(q^2)\neq 0$ for all $q^2$.

When $\phi(q^2)$ has a zero, we have a Fredholm equation of
the third kind. In that
case, both the Fredholm solution, and the Liouville-Neumann
iterations fail.
This is because, effectively, the Fredholm solution involves
integrals of
$\int dq^2 K(q^2,k^2)/\phi(q^2)$, which are ill-defined, whereas
the
Liouville-Neumann solution involves a series with terms of the
form
$\int dk^2 K(q^2,k^2)\phi(k^2)/\phi(q^2)$, which clearly
diverges near the zero of
$\phi$.

In fact it is clear that the solution dramatically changes. When
$\phi(\mu_f^2)=0$, the solution ${\cal F}$ can at best determined
up to a
constant as the equation becomes homogeneous. Furthermore, it is
not in
general possible to remove the pole that the zero of $\phi$
introduces in
${\cal F}$, hence the propagator develops an imaginary part,
which is not
allowed in the $t$-channel.
$\phi(q^2)$ is 1 at $\alpha_S=0$ and steadily
decreases until it reaches a zero value (Note that
$\Delta_1(k^2,q^2)\rightarrow 0$ as $q^2\rightarrow 0$,
hence $\phi$ is finite for all $q^2$).
$\phi$ becomes zero at small $q^2$ first,
and once
$\phi(0)$ has crossed zero, then there will be a zero of $\phi(q^2)$ at some
nonzero $q^2$ for
larger values of $\alpha_S$. Therefore there is a critical value
of
$\alpha_S$ past which $\phi$ is not positive definite.

Physically, one can view the divergence of the
kernel as
the divergence of the effective coupling $\alpha_f(q^2)$ of
Eq.~(\ref{newcoupling}) at some value of $q^2$. This divergence
suggests that
the chiral equation stops having physically relevant solutions,
and hence that
chirality needs to be broken past a certain critical value of the
coupling.
One should really speak of the value of $\alpha_S(\mu)
\Z_R(\mu)$, as one
can always change the value of $\alpha_S$ according to
Eq.~(\ref{RG}).
This interpretation is reinforced when one realizes that
the singularity of the kernel is entirely due to the fact that we
have neglected the terms proportional to $G$. If these are
reintroduced, the
singularity will disappear, and one will keep a solution. Hence,
the
singularity is the place at which the quarks develop a mass.

We can in fact study this for a general gluon propagator, and
hence not limit
ourselves to the solution found in ref.~\cite{us}. We simply need to
assume that
the gluon propagator has a K\"allen-Lehmann representation:
\beq
{{\cal Z}(q^2)\over q^2}=\int d\sigma{\rho(\sigma)\over
q^2+\sigma}
\eeq
$\phi(q^2)$ can
then be written, using the condition ${\cal Z}(\mu_g)=1$:
\begin{eqnarray}
\phi(q^2)&=&\int d\sigma\rho(\sigma)
{\Phi}(\sigma,q^2)\nonumber\\
{\rm with\ }{\Phi}(\sigma,q^2)&=&
\left\{ {\mu_g^2\over \mu_g^2+\sigma} - \ {C_F\ \alpha_S\over
4\pi}
 \int dk^2  \Delta_1(k^2,q^2){1\over k^2+\sigma}\right\}
\label{gsigma}
\end{eqnarray}

\begin{figure}
{\noindent\small Figure 3: Massless quark solutions to the
Dyson-Schwinger equations
exist in the shaded region, the boundary of which is the value of
$\alpha_S$
for which the equation becomes singular for some value of $q^2$,
given the mass squared $\sigma$ which
enters the K\"allen-Lehmann representation of the gluon
propagator. The thick
curve shows the values of $\alpha_S$ at which the effective quark
coupling
(\ref{newcoupling}) becomes infinite. The curves are for
$\eta=1$.}
\end{figure}

We show in Fig.~3 the region in the ($\alpha_S$, $\sigma/\mu_g^2$)
plane in which
${\rm with\ }{\Phi}(\sigma,q^2)$ is positive for all $q^2$. This
means that
if $\rho(\sigma)$ has support in an interval
$[\sigma_0,\sigma_1]$, then there
will be massless solutions if $\alpha_S\leq
\alpha_S^{critical}(\sigma_1)$.

We have plotted Fig.~3 for the function $\eta$ of Eq.~(\ref{dval})
equal to one. It is easy to see that a critical $\alpha_S$ will
exist for a broad range of $\eta$ and that its value can be
calculated from Fig.~3. Indeed, as shown in the Appendix, for
$\eta\left(0,{q^2p^2\over (q^2+p^2)^2}\right)$, we can
write
$\Delta_1=\eta\Delta_1^T + \Delta_1^L$. Unless $\eta$ is such
that the sign of
\beq
I(\eta)=\int dk^2  {[\Delta_1^L(k^2/q^2)+\eta
\Delta_1^T(k^2/q^2)]\over( k^2+\sigma)}
\eeq
changes, there will exist a critical $\alpha_S$.  Its value will
be $\alpha_c(\eta)=\alpha_c(\eta=1) I(\eta=1)/I(\eta)$. We therefore
conclude that there exists a wide range of choices for the
transverse vertex which give rise to chiral symmetry breaking.

It is of course true that even for ${\Phi}$ negative in part of
$[\sigma_0,\sigma_1]$ it is possible to get $\phi>0$, hence the
condition is
sufficient only: in the shaded region, there will be massless
solutions.
It is interesting to note that large-$\sigma$ modes in the
K\"allen-Lehmann
density of the gluon imply chiral symmetry breaking in the quark
sector.
Only for a specific gluon propagator can one
find the exact value of $\alpha_S$ beyond which these solutions
do not exist anymore. This is what we are going to do in the next
section.
\subsection{Chiral solution for a specific gluon propagator}
As explained in Section 2, the gluon propagator (\ref{gluz}) that
we found in
ref.~\cite{us} has the problems linked with the breakdown of
multiplicative
renormalisability. The procedure used here to obtain a consistent ansatz for
the
transverse part of the
vertex can be extended to the gluon case. However, one may hope
that the
effect on gluons will be less dramatic than that on quarks.
Indeed,
we found that imposing the true asymptotic behaviour as
$q^2\rightarrow\infty$
does not appreciably change the behaviour of the propagator at
moderate $q^2$
(compare the solution of the first paper of ref.~{1} with
Eq.~(\ref{gluz})).

Hence we shall use that solution in the quark equation as an
example of what
the chiral solution looks like. We shall assume that both
equations are
renormalised at the same point, and that the two couplings have
the same
value. As we already explained, one should be able to derive a
gluon
propagator for any value of $\alpha_S(\mu)$ as it can always be
changed
according to Eq.~(\ref{gluz}). As our gluonic equation did not
respect
multiplicative renormalisability, we were not able to do so, and
got an
optimum $\alpha_S\approx 1.4$. To see what the effect of the
coupling on the
quark propagator is, we shall vary $\alpha_S$ independently of
${\cal Z}$,
although the two are really correlated.

\begin{figure}
{\small \noindent Figure 4: The solutions for massless quark
propagators that come
from the nonperturbative gluon propagator of ref.~\cite{us}, for
$\alpha_s(mf)=
0.2$ (plain), $0.6$ (dashed), $1.0$ (dot-dashed) and $1.4$
(dashed).}
\end{figure}

We show the result of this exercise in Fig.~4. As the value of
the coupling
grows, the quark propagator is enhanced near the origin. As
$\alpha_S$ grows
further, the propagator experiences oscillations, until the
chiral solution is
lost. It is interesting to note that the value of $\alpha_S$ that
we obtained
for the gluon propagator is very close to the critical value
beyond which the
chiral solution disappears. We also see that the criterion of
Section 5.2 can
in practice work backwards, i.e. it not only predicts when there
will be
chiral solutions, but also when these will disappear.

In order to investigate the sensitivity of the solution to our
choice of the function $\cal D$ in Eq.~(\ref{transverse}), we
consider two possible choices for that function. In Fig.~5, we
show the behaviour of our solution for  for ${\cal D}=q^4+p^4$, instead
of ${\cal D}=(q^2+p^2)^2$
(as in Fig.~3). As expected, the two
solutions have the same high-$q^2$ behaviour and are equal at
$q^2=\mu_g^2$.
\begin{figure}
{\small \noindent Figure 5: Same as Fig.~4, but for with a transverse vertex
multiplied by $(q^2+p^2)^2\over q^4+p^4$.}
\end{figure}
\noindent We see that the low-momentum behaviours vary
by a modest amount, and that the solutions only differ in the region
$q^2\approx \mu_f^2$, and only when the integrand oscillates a lot.
This is where we do not trust our approximation for the angular integrals
and hence the result is stable where our approximations hold.

\vskip 16pt
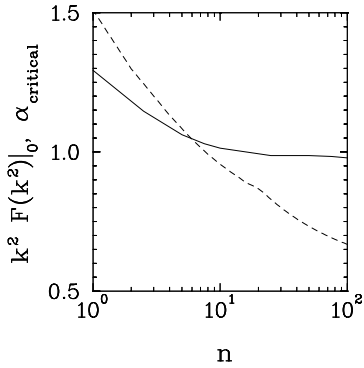
{\small \noindent Figure 6: The dashed curve shows the value of $\alpha_c$ for
various functions $\eta$ (see text), and the plain curve shows the value of
the intercept of the propagator for $\alpha_S=0.75\ \alpha_c$.}

Finally, Fig.~6 illustrates the insensitivity of our results to the choice
of transverse vertex. We consider the functions
$\eta=\left[(q^4+p^4)/(q^2+p^2)^2\right]^n$ and plot the results in terms of
$n$.
We show the variation of the critical value of $\alpha_S$, which changes by a
factor 2 when n changes by a factor 100. We also show the value of the
intercept at the origin of the propagator calculated at $\alpha_S=0.75
\alpha_c$, and which hardly changes with $n$. This clearly illustrates that
the results we have obtained hold
for a wide class of transverse vertices.
\section{Conclusion}
We have shown that the transverse part of the vertex plays an
essential role
in the quark DS equation in the axial gauge. It restores
multiplicative
renormalisability, allows chiral symmetry breaking and provides
solutions
which match with perturbation theory at large $q^2$. Hence we
have obtained an
equation that possesses all the properties that are required in
QCD.

The solution that we obtain in the chiral phase
has a pole at $q^2=0$ despite the fact that we expect quarks to
be confined.
This contrasts with the results of a similar analysis carried out
in ref.~\cite{mike1} for the gluon propagator of ref.~\cite{bbz}
in which
it was shown that for the {\it confining} propagator of
ref.~\cite{bbz} the
quark propagator does not have a singular behaviour at $q^2=0$
and can
therefore be considered to be confined.
The present formalism can be used to solve for the chirality
breaking
solution, as we shall explain in a future publication
\cite{uslater}.
It is an interesting
question to see whether these solutions are confined, and hence
whether
confinement and chiral symmetry breaking are related.

It has not so far been possible to obtain a solution to the
(modified) DS equations for the gluon propagator for time-like
momenta. This is because we expect that in this regime the
propagator
develops an imaginary part with cuts corresponding to the
thresholds
for glueball production etc. and consequently such a solution
involves
coupled non-linear equations for the real and imaginary parts of
$\Z_R(k^2)$.
Therefore the corresponding solution
for the quark propagator we have obtained is only valid for
quarks propagating with space-like momenta (in the t-channel). It
is
tempting to try to analytically continue the solution obtained
into
time-like momenta by fitting the solution obtained to known
analytic
functions and possibly then using the DS equation for the
quark propagator to extract information about the gluon
propagator with
time-like momenta.  Unfortunately this has not been possible. We
have
obtained several very accurate fits (errors nowhere worse than
2\% ) to
the curve shown in Fig.~4, using different parametrisations
inside
various analytic functions. Whereas these all fit the quark
propagator
remarkably well in the fit region they give wildly different
projected behaviours for the quark propagator for time-like
momenta. We
therefore have to accept for the moment that the important
problem of
describing the exchange of soft gluons and quarks in the
s-channel
remains unsolved.

As we explained earlier, the equation for the gluon propagator
also will have
to be modified to incorporate a transverse vertex and recover
multiplicative
renormalisability. Furthermore,
as has been pointed out in ref.~\cite{mike2} the DS equations for
the quark and gluon propagators are coupled and after having
obtained the solution for the
quark propagator one must check
the
assumption that one may neglect quark loops in the equation for
the gluon propagator.
In ref.~\cite{mike2} it was shown that for the
gluon
propagator of ref.~\cite{bbz} this was not the case and the gluon
propagator had to be modified accordingly. A similar analysis
should
also be carried out for the gluon and quark propagators discussed
here.

\vspace{1in}
\noindent {\Large {\bf Acknowledgements:}} \\
JRC and DAR wish to thank the Physics Departments at McGill
University and at The
University, Southampton, to have enabled them to visit each other
and to carry
out this work. This work was supported by NSERC (Canada), Les
Fonds FCAR
(Qu\'ebec) and PPARC (United Kingdom).

\vfill\break
\section[*]{Appendix: Kernels}
 The kernels entering
Eq.~(\ref{equation}) are given by the following expressions,
where we define  $\rho=k^2/q^2$.

For the BZ equation (\ref{feqa}):
\beq {\Delta}_1(\rho) \ = \ {\rho\over 4} -
     \theta (\rho - 4 ) \sqrt{1-{4\over\rho}}
  \left(\frac{\rho}{4 }+\frac{1}{2 } \right)  \eeq
and
\begin{eqnarray}
 {\Delta}_2(\rho) \ & = &  \ \theta (1-\rho)\frac{\rho
(\rho+1)}{4}
    + \theta (\rho-1)
   \left( \frac{3}{2}(1+{1\over\rho})
     -\frac{\rho(\rho+1)}{4} \right) \nonumber  \\ & &
     +\theta (\rho - 4) (1+\rho)
    \sqrt{1-{1\over\rho}}\left( \frac{\rho}{4 }
 +  \frac{1}{2} \right)
\end{eqnarray}

For Eq.~(\ref{feq}), and for $\eta$ a function of ${q^2p^2\over
(q^2+p^2)^2}\approx {1+\rho\over (2+\rho)^2}$ only:
\begin{eqnarray}
   \Delta_1(\rho) &=&
    {\eta\over 4\rho} \left[ - \rho- 8 +
{  \rho^2 + 16 + 2\rho \over\sqrt{\rho^2+4}}
\right]\nonumber\\
       &+& {1\over 4}\left[ {- \rho^2 + 2\, \, \rho + 8\over
\sqrt{\rho^2-4\, \rho}}\, + \rho\right]\nonumber\\
      &+& \theta(4-\rho) \, {1\over 4}
{\rho^2 - 2\, \rho - 8 \over \sqrt{\rho^2-4\, \rho}}
\end{eqnarray}

 \begin{eqnarray}
 \Delta_2(\rho) &=&
{ \eta\over 4\rho} \,  \left\{ 9\rho +  \rho^2
+ 8
- {1\over \sqrt{\rho^2+4}} [ \rho^3 + \rho^2 + 18\rho
+ 16]\right\}\nonumber\\
 &+& \theta(\rho-1) \,   {3\over 2 \rho}\nonumber\\
 &+& \theta(1-\rho) \,   {- 3 + \rho^2 + \rho\over 2}\nonumber\\
 &+& \theta(4-\rho) \,
 {  - \rho^3    + \rho^2    + 10\,\rho
   +8\over4\sqrt{\rho^2-4\, \rho}}
          \nonumber\\
  &+&
{6 - \rho^2
- \rho\over 4}
+{ \rho^3\, -  \rho^2 - 10\rho
- 8\over 4\sqrt{\rho^2-4\, \rho}}
\end{eqnarray}
\vfill\break

\newpage

\end{document}